
%
%
\input phyzzx
\hoffset=0.2truein
\hsize=6truein
\voffset=0.1truein
\def\TITLEPAGE{\frontpagetrue}
\def\PUPT#1{\hbox to\hsize{\tenpoint \baselineskip=12pt
        \hfil\vtop{
        \hbox{\strut PUPT-95-#1}
}}}
\def\PRINCETON{
\centerline{Physics Department, Princeton University}
\centerline{Princeton, New Jersey 08544}}
\def\TITLE#1{\vskip .0in \centerline{\fourteenpoint #1}}
\def\AUTHOR#1{\vskip .1in \centerline{#1}}

\def\ABSTRACT#1{\vskip .1in \vfil \centerline{\twelvepoint
\bf Abstract} #1 \vfil}
\def\ENDTITLEPAGE{\vfil\eject\pageno=1}
\hfuzz=5pt
\tolerance=10000
\TITLEPAGE

\PUPT{1518}

\vskip 1in

\TITLE{Open Inflation with Arbitrary False Vacuum Mass}

\vskip 1in

\AUTHOR{Martin Bucher and Neil Turok}

\vskip 5pt

\PRINCETON

\ABSTRACT{
We calculate the power spectrum of
adiabatic density perturbations in an open inflationary
model in which inflation occurs in two stages. First
an epoch of old inflation creates a large, smooth universe,
solving the horizon and homogeneity problems. Then an open universe
emerges
through the nucleation of a single bubble,
with constant density  hypersurfaces inside the bubble having
constant negative spatial curvature. An epoch of `slow roll'
inflation, shortened to give $\Omega _0<1$ today, occurs within
the bubble, which contains our entire observable
universe. In this paper we compute the resulting
density perturbations in the same `new thin wall' approximation
used in a previous paper, but for an arbitrary positive
mass of the inflaton field in the false vacuum.
}

\rightline{January 1995}

\ENDTITLEPAGE

\eject

\REF\guth{A. Guth, ``Inflationary Universe: A Possible Solution to the Horizon
and Flatness Problems," Phys. Rev {\bf D23,} 347 (1981).}

\REF\ninf{A. Linde, ``A New Inflationary Universe Scenario: A Possible
Solution of the Horizon, Flatness, Homogeneity, Isotropy, and
Primordial Monopole Problems," Phys. Lett. {\bf 108B,} 389 (1982);
A. Albrecht and P. Steinhardt, ``Cosmology for Grand Unified Theories
with Radiatively Induced Symmetry Breaking," Phys. Rev. Lett. {\bf 48,} 1220
(1982).}

\REF\lindec{A. Linde, ``Chaotic Inflation," Phys. Lett.
{\bf 129B,} 177 (1983).}

\REF\cd{S. Coleman and F. De Luccia, ``Gravitational Effects on and of
Vacuum Decay," Phys. Rev. {\bf D21,} 3305 (1980).}

\REF\gott{J.R. Gott, III, ``Creation of Open
Universes from de Sitter Space,"
Nature {\bf 295,} 304 (1982); J.R. Gott and
T. Statler, ``Constraints on the
Formation of Bubble Universes," Phys. Lett. {\bf 136B,} 157 (1984);
J.R. Gott, ``Conditions for the Formation of Bubble Universes," in
E.W. Kolb et al., Eds., {\it Inner Space/Outer Space,} (Chicago:
U. of Chicago Press, 1986).}

\REF\bgt{M. Bucher, A.S. Goldhaber, and N.Turok, ``An Open
Universe From Inflation," Princeton Preprint
(hep-ph 94-11206) (1994).}

\REF\sasaki{ M. Sasaki, T. Tanaka, K. Yamamoto, and J. Yokoyama,
``Quantum State During and After Nucleation of an
$O(4)$ Symmetric Bubble,";  Prog. Theor. Phys. {\bf 90,}
1019 (1993); M. Sasaki, T. Tanaka, K. Yamamoto, and J. Yokoyama,
``Quantum State Inside a Vacuum Bubble and Creation of
an Open Universe," Phys. Lett. {\bf B317,} 510 (1993);
T. Tanaka and M. Sasaki,
``Quantum State During and After O(4) Symmetric Bubble
Nucleation with Gravitational Effects", Phys. Rev. {\bf D50,}
6444 (1994);
K. Yamamoto, T. Tanaka and M. Sasaki,
``Particle Creation through Bubble Nucleation and Quantum Field Theory
in the Milne Universe," preprint (1994).}

\REF\birrell{ N. Birrell and P. Davies, {\it Quantum Fields in
Curved Space,} (Cambridge, Cambridge U. Press, 1982) and
references therein.}

\REF\erdelyi{A. Erdelyi et al.,
{\it Higher Transcendental Functions,}  Vol. 1, (New York:
McGraw-Hill) (1953).}

\REF\whipp{W.N. Bailey, {\it Generalised Hypergeometric Series,}
(London, Cambridge University Press) (1935).}

\REF\lyth{D. Lyth and E. Stewart, ``Inflationary Density Perturbations with
$\Omega <1,$" Phys. Lett. {\bf B252,} 336 (1990).}

\REF\ratra{B. Ratra and P.J.E. Peebles, ``CDM Cosmogony in an Open Universe,''
Ap. J. {\bf 432}, L5, (1994).}

\REF\ratrb{B. Ratra and P.J.E. Peebles, ``Inflation in an Open Universe,''
preprint PUPT-1444 (1994).}

\REF\kamion{ M. Kamionkowski,
B. Ratra, D. Spergel, and N. Sugiyama,
``CBR Anisotropy in an Open Inflation, CDM Cosmogony'',
Ap. J. {\bf 434}, L1 (1994).}

\REF\gorski{
K. Gorski, B. Ratra, N. Sugiyama and A. Banday,
``COBE-DMR-Normalized Open Inflation, CDM Cosmogony,''
Princeton preprint  PUPT-1513, CfPA-Th-94-61, UTAP-194,
astro-ph 9502034.}

\REF\hmoss{S. Hawking and I. Moss, ``Supercooled Phase
Transitions in the Very Early Universe," Phys. Lett. {\bf 110B,}
35 (1982).}

\REF\jensen{L. Jensen and P. Steinhardt, ``Bubble Nucleation and the
Coleman-Weinberg Model," Nucl. Phys. {\bf B237,} 176 (1984).}

\REF\newa{A. Starobinsky in
H.J. de Vega and N. Sanchez, Eds.,
{\it Current Topics in Field Theory,
Quantum Gravity and Strings,} Lecture Notes in Physics {\bf 206,}
(Heidelberg: Springer)(1986);
A. Goncharov and A. Linde, ``Tunneling in Expanding Universe:
Euclidean and Hamiltonian Approaches," Sov. J. Part. Nucl.
{\bf 17,} 369 (1986);
A. Linde, ``Stochastic Approach to Tunneling and Baby Universe
Formation," Nucl. Phys. {\bf B372,} 421 (1992).  }

\REF\bucher{M. Bucher and N. Turok, in preparation.}

\REF\yst{K. Yamamoto, M. Sasaki, and T. Tanaka, ``Large
Angle CMB Anisotropy in an Open Universe in the
One-Bubble Scenario," Kyoto preprint KUNS 1309,
astro-ph 95-1109 (1995).}

\def\Box{\mathord{\dalemb{5.9}{6}\hbox{\hskip1pt}}}
\def\dalemb#1#2{{\vbox{\hrule height.#2pt
	\hbox{\vrule width.#2pt height#1pt \kern#1pt \vrule width.#2pt}
	\hrule height.#2pt}}}

\def\Box{\mathord{\dalemb{5.9}{6}\hbox{\hskip1pt}}}
\def\dalemb#1#2{{\vbox{\hrule height.#2pt
	\hbox{\vrule width.#2pt height#1pt \kern#1pt \vrule width.#2pt}
	\hrule height.#2pt}}}
\def\sech{ {~\rm sech ~ }}

\def\gtorder{\mathrel{\raise.3ex\hbox{$>$}\mkern-14mu
             \lower0.6ex\hbox{$\sim$}}}
\def\ltorder{\mathrel{\raise.3ex\hbox{$<$}\mkern-14mu
             \lower0.6ex\hbox{$\sim$}}}

\def\overleftrightarrow#1{\vbox{\ialign{##\crcr
$\leftrightarrow$\crcr\noalign{\kern-1pt\nointerlineskip}
$\hfil\displaystyle{#1}\hfil$\crcr}}}

\chapter{Introduction}

\def\np{{\nu ^\prime}}
\def\overleftarrow#1{\vbox{\ialign{##\crcr
$\leftarrow$\crcr\noalign{\kern-1pt\nointerlineskip}
$\hfil\displaystyle{#1}\hfil$\crcr}}}
\def\overrightarrow#1{\vbox{\ialign{##\crcr
$\rightarrow$\crcr\noalign{\kern-1pt\nointerlineskip}
$\hfil\displaystyle{#1}\hfil$\crcr}}}

If the current density parameter $\Omega _0$ is smaller than unity
as some observations suggest,
then we live in an open universe, described to a first approximation
by a
Friedmann-Robertson-Walker (FRW) metric of the form
$$ds^2=-dt^2+a^2(t)\cdot [d\xi ^2+\sinh ^2[\xi ]d\Omega _{(2)}^2].
\eqn\iaa$$
Slices of constant cosmic time are maximally symmetric three
dimensional manifolds of constant negative spatial curvature.
The symmetry group of such a universe is
$SO(3,1),$ with `boosts' corresponding to
what we commonly regard as spatial translations.

Most inflationary models\refmark{\guth ,\ninf ,\lindec }
predict a value of
$\Omega _0$ extremely
close to unity. However, this is not a necessary consequence
of inflation. As noted by Coleman and de Luccia\refmark{\cd }
and by Gott, \refmark{\gott }
when a bubble nucleates in de Sitter space, inside the forward
light cone of the materialization center, spatial
hypersurfaces on which the
scalar field is constant are spaces of constant negative
spatial curvature. In other words, the bubble contains an
expanding open FRW universe. Consequently, one can create an open
universe from inflation through a two-stage
process.\refmark{\gott ,\bgt }  During an initial
epoch of {\it old} inflation the inflaton field is stuck in
a false vacuum. In this epoch the smoothness and horizon
problems are solved; whatever inhomogeneities may have existed
prior to inflation are erased. Then {\it old} inflation
is exited through the nucleation of a single bubble. Instead
of tunneling directly to the true vacuum, the inflaton field
tunnels onto a `slow roll' potential, and a shortened epoch
of {\it new} inflation occurs inside the bubble. By {\it new}
inflation, we mean here slow-roll inflation and do not
refer to the way in which inflation began. For our purposes
it is sufficient to assume that a sufficiently large volume
became stuck in the false vacuum. This may have happened
{\it chaotically,}\refmark{\lindec } or perhaps in some other way.
Formally $\Omega $ is exactly zero
on the forward light cone of the materialization center,
and $\Omega$ flows toward one during the epoch of {\it new} inflation
inside the bubble until that era ends at reheating.
In the subsequent evolution $\Omega $ flows away from unity.
Its present value,  $\Omega_0$,  is
determined by the total expansion factor during the
{\it new} inflationary epoch and by the final reheat temperature.
The usual fine tuning argument against $\Omega _0\ne 1$ is not
valid here, because it is not $[\Omega^{-1} -1]$ at reheating but
rather ${\rm ln}[\Omega^{-1} -1]$ at reheating that is proportional
to the length of the new inflationary epoch. Thus to obtain
interesting values for $\Omega_0$, one has to tune not
a very small number, but rather the logarithm of a very small number.
Numerically, this turns out to be a rather mild requirement.
\refmark{\bgt }

Prior to bubble nucleation the geometry of spacetime is that
of pure de Sitter space, with $H^2=(8\pi G/3)V[\phi _{fv}]$
where $\phi _{fv}$ is the expectation value of the inflaton
field in the false vacuum. To exploit the $SO(3,1)$ symmetry
of the expanding bubble solution, it is advantageous to
work in hyperbolic coordinates, which divide maximally extended
de Sitter space into five coordinate patches, shown in Fig. 1,
only two of which shall concern us here.

The line element for region I is given in eqn. \iaa , and for
de Sitter space $a(t)=H^{-1}\sinh [Ht].$ The line element
for region II is
$$ ds^2=d\sigma ^2+b^2(\sigma )\cdot \left[
-d\tau ^2+\cosh ^2[\tau ]d\Omega _{(2)}^2\right] ,\eqn\iab $$
where for de Sitter space $b(\sigma )=H^{-1}\sin [H\sigma ],$ with
$0<H\sigma <\pi .$ [Regions III, IV, and V have line elements
of the form given in eq. \iaa .]

Hyperbolic coordinates are useful for
describing the expanding bubble solution because the inflaton
field (in the background solution) is constant on slices of
constant $t$ (in region I) and on slices of constant $\sigma $
(in region II). In Fig. 2 is sketched a bubble nucleation event,
the solid lines indicating the surfaces on which the scalar
field is constant. The horizontal dashed line separates
the Euclidean (classically forbidden) region below
from the Lorentzian (classically allowed) region above.

To compute the spectrum of density perturbations produced
by quantum fluctuations of the inflation field, one has to
evolve the mode  functions from the external de Sitter
space across the bubble wall and into the bubble's interior.
In a previous paper, \refmark{\bgt } we performed
this calculation in three stages. First we expressed
the `Bunch--Davies' vacuum modes (the natural vacuum
modes for de Sitter space, see below) in hyperbolic
coordinates. We then matched these modes across
the bubble wall, taken to be very thin. Finally
we included the coupling to gravity and computed the
spectrum of density perturbations including gravity
in the interior of the bubble. Because of the technical
difficulty of the calculations, we restricted ourselves
to a special case, where the mass $m^2$ of the inflaton
field in the false vacuum equals $2H^2$, with $H$ the
Hubble constant during old inflation.
As we emphasized there, this was an assumption
without physical basis, solely made to simplify the
computation. In this paper we generalize the result
to arbitrary positive
$m^2/H^2.$ For those parts of the computation that are
identical, we cite the results from ref. \bgt .

Some of these issues have been investigated
using a technique involving analytic continuation of Euclidean modes
in a series of recent paper.
\refmark{\sasaki} (See Note Added).

The organization of the paper is the following. In
section II we discuss initial conditions. The Bunch--Davies
vacuum for a massive scalar field is expanded
in terms of region II hyperbolic modes. In section 3,
we continue these modes into the bubble coupling to
the scalar component of linearized gravity and
calculate the power spectrum. In section IV we
present some concluding remarks.

\chapter{Initial Conditions}

Initially, prior to bubble materialization, the fluctuations
of the inflaton field about the false vacuum may be regarded
as a free scalar field of mass $m^2=V^{\prime \prime }[\phi _{fv}].$
The so-called Bunch--Davies vacuum is the natural initial
quantum state for these fluctuations for the following reason.
Let us imagine there is some state prior to the onset of
old inflation. This state may be quite inhomogeneous,
but if it is to have a finite
(renormalized) energy density, the very short wavelength field modes
must be taken to be in their ground state.
At very short distances the spacetime
approaches Minkowski spacetime, and the effects of spacetime
curvature can be ignored (at least in the naive approach to
quantizing fields in curved backgrounds that we shall follow
here---see, for example, ref. \birrell ).
Ground state here means Minkowski space
vacuum. As inflation begins, the
co-moving wavelengths of all field modes are exponentially stretched.
After a certain amount of {\it old} inflation (the same amount needed
to make the universe homogeneous and isotropic) the only modes
of interest are those which were exponentially
far within the horizon when inflation began.  It is
natural to assume these modes are in the state
forced upon them by the finite initial density constraint
(i.e., the state corresponding to the
{\it Minkowski space vacuum} at early times).
This is the so-called Bunch--Davies vacuum. (For a discussion,
see ref. \birrell .) Needless to say, we should
be reluctant to drop this assumption, because without it
the mechanism of quantum-fluctuation generated
perturbations would be likely to lose any  predictive power.

We shall imagine that enough old inflation occurred
to produce a homogeneous and isotropic universe, and
to `drive' the scalar field modes of interest into
the Bunch-Davies vacuum, via the constraint explained above.
Our first task then is to express the Bunch--Davies vacuum in terms
of region II hyperbolic modes, so that we can continue these
modes into region I and after coupling to the scalar
component of gravity calculate the power spectrum from open
inflation. The wave equation for the inflaton field
in terms of the region II hyperbolic coordinates
is
$$\eqalign{&-\bigl[ \Box +m^2(\sigma )\bigr] \cdot
\phi (\sigma , \tau ,\theta ,\phi )\cr
&~~=\left[ {\partial ^2\over \partial \sigma ^2}
+3\cot [\sigma ] {\partial \over \partial \sigma }
-{1\over \sin ^2[\sigma ]}\cdot \left(
{\partial ^2\over \partial \tau ^2}+2\tanh [\tau ]
{\partial \over \partial \tau } +{{\bf L}^2\over \cosh ^2[\tau ]}
\right) -m^2(\sigma )
\right] \cr
&~~~~~\times \phi (\sigma , \tau ,\theta ,\phi )=0\cr }
\eqn\ica $$
where we work in units with $H=1$ and where ${\bf L}^2$ is
the usual angular momentum operator. Using the `new thin wall
approximation,' discussed in detail in ref. [\bgt ], we set
$m^2$ equal to $V^{\prime \prime }[\phi _{fv}]$ everywhere
in region II and equal to zero everywhere in region I.
To compute the power
spectrum, it is sufficient to consider only the $s$-wave
sector, so we set ${\bf L}^2=0.$ Eqn. \ica ~ can be solved
by separation of variables, where
$$ \phi ^{(\pm )}(\sigma ,\tau ;\zeta )=
S(\sigma )\cdot {e^{\mp i\vert \zeta \vert \tau }\over \cosh [\tau ]},
\eqn\icb$$
where $S(\sigma )$ satisfies
$$
S^{\prime \prime }(\sigma )+3\cot [\sigma ]S^{\prime }(\sigma )
+\left[ {\zeta ^2\over \sin ^2[\sigma ]}-m^2
\right] S(\sigma )=0.\eqn\icc$$
After the change of dependent variable
$S(\sigma )=F(\sigma )/\sin [\sigma ]$
and the  change of independent variable
$x=\cos [\sigma ],$
eqn. \icc ~ becomes the Legendre equation and
$$S(\sigma ;\zeta )= { {\rm P}^{i\zeta }_\np (\cos [\sigma ])
\over \sin [\sigma ] }\eqn\icd$$
where $\np =\sqrt{{9\over 4}-m^2}-{1\over 2}.$ [There
are two linearly independent solutions, one with
${\rm P}^{i\zeta }_\np $ and the other with $ {\rm P}^{-i\zeta }_\np .$]
We define as usual (see, e.g., ref. \erdelyi , p. 143)
$$
{\rm P}^{i\zeta }_\np (x)={1\over \Gamma (1-i\zeta )}
\left( {1+x\over 1-x}
\right) ^{i\zeta /2}\cdot
{}_2F_1\bigl( -\np , \np +1; 1-i\zeta ; {1-x\over 2}
\bigr) .
\eqn\ice$$
In terms of $u,$ where $\tanh [u]=x=\cos [\sigma ],$
$$S_\zeta =
{1\over \sech [u]}\cdot {1\over \Gamma (1-i\zeta )}
\cdot e^{+i\zeta u}\cdot
{}_2F_1\bigl( -\np , \np +1; 1-i\zeta ; {1\over 1+e^{2u}}
\bigr) .\eqn\icf$$
{}From the self-adjointness properties of the Legendre
equation, it follows that
$$\eqalign{\Gamma (1-\zeta )~\Gamma (1-\zeta ')~&
\int _{-\infty }^{+\infty }du ~{\rm P}^{i\zeta }_\np (\tanh [u])~
{\rm P}^{i\zeta '}_\np (\tanh [u])\cr
&= C_1(\zeta )\cdot \delta (\zeta +\zeta ')
+C_2(\zeta )\cdot \delta (\zeta -\zeta ')\cr }\eqn\ich
$$
where the functions $C_1(\zeta )$ and $C_2(\zeta )$ are
to be determined from the
asymptotic behavior of ${\rm P}^{i\zeta }_\np $ as
$u\to +\infty $ and $u\to -\infty .$ Note that since $\nu'$
is either real or of the form $-{1\over 2} +i\gamma$ with
$\gamma$ real, it follows that
$[{\rm P}^{i\zeta }_\np (\tanh [u])]^* =
{\rm P}^{-i\zeta }_\np (\tanh [u]).$
Clearly, as
$u\to +\infty ,$
$$ \Gamma (1-i\zeta )~{\rm P}^{i\zeta }_\np \approx  e^{+i\zeta u}.
\eqn\icg$$

{}From the relation
$$\eqalign{
{}_2F_1\bigl( -\np , &\np +1; 1-i\zeta ; 1-w \bigr) \cr
&= {\Gamma (1-i\zeta ) \Gamma (-i\zeta )\over
\Gamma (1-i\zeta +\np ) \Gamma (-i\zeta -\np )}~~
{}_2F_1\bigl( -\np , \np +1; 1+i\zeta ; w \bigr) \cr
&~~+~w^{-i\zeta }~~{\Gamma (1-i\zeta ) \Gamma (+i\zeta )\over
\Gamma (1+\np )\Gamma (-\np )}~~
{}_2F_1\bigl( -\np , \np +1; 1-i\zeta ; w \bigr) ,\cr }
\eqn\ici$$
it follows that as $u\to -\infty ,$
$$\eqalign{\Gamma (1-i\zeta )
{\rm P}^{i\zeta }_\np \approx & {\Gamma (1-i\zeta )\Gamma (-i\zeta )\over
\Gamma (1-i\zeta +\np ) \Gamma (-i\zeta -\np )}~ e^{+i\zeta u}\cr
&+ {\Gamma (1-i\zeta )\Gamma (+i\zeta )\over
\Gamma (1+\np )\Gamma (-\np )}~e^{-i\zeta u} .\cr }
\eqn\icj$$
Consequently, using
$$\int _0^\infty du~e^{i\zeta u}=\pi \cdot \delta (\zeta )+
i{\rm P.P.}\left( {1\over \zeta} \right)
\eqn\icjj$$
where P.P. indicates the principal part, one obtains
$$\eqalign{ C_1(\zeta )&=\pi \cdot \biggl[
1+ {\Gamma (1-i\zeta )\Gamma (1+i\zeta )
\Gamma (-i\zeta )\Gamma (+i\zeta )\over
\Gamma (1-i\zeta +\np )\Gamma (1+i\zeta +\np )
\Gamma (-i\zeta -\np ) \Gamma (+i\zeta -\np )} \cr
& +{\Gamma (1-i\zeta )\Gamma (1+i\zeta )
\Gamma (-i\zeta )\Gamma (+i\zeta )\over
\Gamma ^2(1+\np )\Gamma ^2(-\np )}
\biggr] \cr
&=(2\pi )\cdot \left[ 1+{\sin ^2[\pi \np ]\over \sinh ^2[\pi \zeta ]}
\right] ,\cr }\eqn\ick$$
and similarly
$$\eqalign{
C_2(\zeta )&=\pi \cdot
{2\Gamma (1-i\zeta )\Gamma (1-i\zeta )\Gamma (-i\zeta )\Gamma (+i\zeta )
\over \Gamma (1+\np )\Gamma (-\np )
\Gamma (1-i\zeta +\np ) \Gamma (-i\zeta -\np )	}\cr
&=(2\pi )\cdot {\sin [\pi \np ]\over \sinh ^2[\pi \zeta ]}
\cdot {\Gamma (+i\zeta -\np )\over \Gamma (-i\zeta -\np )}\cdot
{\Gamma (1-i\zeta )\over \Gamma (1+i\zeta )}\cr
&~~~\times \biggl[ \cosh [\pi \zeta ] \sin [\pi \np ]-i
\sinh [\pi \zeta ]\cos [\pi \np ] \biggr] .\cr }\eqn\icky$$

\def\F{{\cal F}}

We wish to define spatial mode functions $F^{i\zeta }$ that
are linear combinations of ${\rm P}^{i\zeta }_\np $ and
${\rm P}^{-i\zeta }_\np $ chosen so that
$$
\int _{-\infty }^{+\infty }du~F^{i\zeta }~F^{i\zeta '}
={1\over 8\pi \vert \zeta \vert }\cdot \delta (\zeta +\zeta ').
\eqn\icl$$

The functions $C_1(\zeta )$ and $C_2(\zeta )$ have the form
$$\eqalign{
C_1(\zeta )&=(2\pi )\cdot \cosh ^2[\bar \xi (\zeta )],\cr
C_2(\zeta )&=(2\pi )\cdot \cosh [\bar \xi (\zeta )]
\sinh [\bar \xi (\zeta )]e^{i\bar \varphi (\zeta )} \cr
}\eqn\icla$$
where $\bar \xi (\zeta )$ and $\bar \varphi (\zeta )$ are
real.  Therefore, we may choose $F^{i\zeta }$ according to
$$\eqalign{ F^{+i\zeta }&={1\over 4\pi \sqrt{\vert \zeta \vert
\cosh [\bar \xi ]}}\cr
&\times \biggl[ \cosh [\bar \xi /2]~\Gamma (1-i\zeta )~{\rm P}^{+i\zeta }_\np
-e^{i\bar \varphi } ~\sinh [\bar \xi /2]~
\Gamma (1+i\zeta )~{\rm P}^{-i\zeta }_\np \biggr] ,\cr }
\eqn\icm$$
which in terms of $C_1(\zeta )$ and $C_2(\zeta )$ may be
rewritten as
$$\eqalign{ F^{+i\zeta }={1\over 4\pi \sqrt{\vert \zeta \vert }}
&\times \Biggl[ \sqrt{
1+\sqrt{1-\vert C_2\vert ^2/C_1^2}\over 2}~
\Gamma (1-i\zeta ){\rm P}^{+i\zeta }_\np \cr
&~~~~-{C_2\over \vert C_2\vert } \sqrt{
1-\sqrt{1-\vert C_2\vert ^2/C_1^2}\over 2}~
\Gamma (1+i\zeta ){\rm P}^{-i\zeta }_\np \Biggr] .\cr }
\eqn\icma$$
It follows the creation and annihilation operators associated
with the modes
$$\F _\zeta ^{(\pm )}=F_\zeta (u)\cdot {e^{\mp i\vert \zeta \vert \tau }
\over \cosh [\tau ]}\eqn\icnn$$
where
$$\hat \phi (\xi ,\tau )=\int _{-\infty }^{+\infty }d\zeta ~
\biggl[ \F _\zeta ^{(+)}\hat a^{(+)}(\zeta )
+\F _\zeta ^{(-)}\hat a^{(-)}(\zeta )\biggr] \eqn\icnna$$
obey the usual commutation relations
$$\eqalign{
[\hat a^{(+)}(\zeta ), \hat a^{(-)}(\zeta ')]&=\delta (\zeta -\zeta '),\cr
[\hat a^{(+)}(\zeta ), \hat a^{(+)}(\zeta ')]&=0,\cr
[\hat a^{(-)}(\zeta ), \hat a^{(-)}(\zeta ')]&=0.\cr }
\eqn\ico$$

Although these modes have the correct commutation relations,
they are not useful for calculating expectation values
with respect to the Bunch--Davies vacuum because the operators
$\hat a^{(+)}(\zeta )$ do not annihilate the Bunch--Davies
vacuum. The Bunch--Davies vacuum is related to the vacuum defined
by the annihilation operators by a Bogolubov transformation,
which we shall now calculate.

We form `positive frequency' modes (with respect to
the Bunch--Davies vacuum) by considering linear
combinations of the form
$$ f^{(+)}_\zeta +c_\zeta ~f^{(-)}_\zeta
\eqn\asatz$$
where the coefficients $c_\zeta $ are to be determined.

We may determine the coefficient $c_\zeta ,$ and verify
the validity of the ansatz \asatz ~ as well, by requiring
that products of the form
$$
(f^{(+)}_\zeta +c_\zeta ~f^{(-)}_\zeta ,p)
\eqn\cond
$$
to vanish where $p$ is a Bunch-Davies positive frequency mode,
and where we define
$$ (u, v)=(-i)\int _\Sigma d\Sigma ^\mu ~
\biggl\{ u(X)[\partial _\mu v(X)]-
[\partial _\mu u(X)] v(X) \biggr\} \eqn\jje $$
where $\Sigma $ is a Cauchy surface with unit normal $n^\mu$, and
$d \Sigma^\mu = d\Sigma n^\mu$ with $d\Sigma$ the volume element
on $\Sigma .$ If the product in eqn. \cond ~ vanishes for all $p,$
then $f^{(+)}_\zeta +c_\zeta ~f^{(-)}_\zeta $ is a
positive frequency mode.

The Bunch-Davies positive frequency modes in terms of closed
coordinates are known. They are
$$\phi_k^{BD}={\sin [k\sigma ]\over \sin [\sigma ]}\cdot T_k(\eta )
\eqn\cmde$$
where
$$T_k(\eta )=\sin ^{1\over 2}[\eta ]\cdot \biggl[
{\rm P}^\nu _{k-{1\over 2}}(-\cos [-\eta ])-{2i\over \pi }
{\rm Q}^\nu _{k-{1\over 2}}(-\cos [-\eta ])
\biggr] \eqn\tdep$$
where $\tanh [\eta /2]=e^t.$ Here $k$ is a positive integer.
$\eta ={\pi \over 2}$
corresponds to $\tau =0$ and at $\tau =0$ one
has $\partial _{\hat t}=\partial _\eta .$

The condition
$$\eqalign{
&(f^{(+)}_\zeta +c_\zeta ~f^{(-)}_\zeta ,\phi^{BD}_k)\cr
&~~~=(-i4\pi )\int _0^\pi d\sigma ~\sin ^2[\sigma ]~
\left[
{{\rm P}^{i\zeta }_\np (\cos [\sigma ]) \over \sin [\sigma ]}\cdot
{ e^{-i\vert \zeta \vert \tau } +
c_\zeta ~e^{+i\vert \zeta \vert \tau }
\over \cosh [\tau ]}
\right] \cr &~~~~~~~~~\times
\left[
{\overrightarrow{\partial } \over \partial \eta }
-{1\over \sin [\sigma ]}
{\overleftarrow{\partial } \over \partial \tau }
\right] \times
\left[ {\sin [k\sigma ]\over \sin [\sigma ]}\cdot T_k(\eta ={\pi \over 2})
\right] =0\cr }\eqn\bba$$
is equivalent to
$$\int _0^\pi d\sigma ~{\rm P}^{i\zeta }_\np (\cos [\sigma ])~
\sin [k\sigma ]~
\left[ { T^\prime _k(\eta ={\pi \over 2})
\over  T_k(\eta ={\pi \over 2})}
+{i\vert \zeta \vert \over \sin [\sigma ]}
\left( {1-c_\zeta \over 1+c_\zeta }\right)
\right] =0.\eqn\bbb$$
We thus obtain an infinite number of equations, and the
$k$th equation is solved by
$$
{1-c_\zeta \over 1+c_\zeta }={i\over \vert \zeta \vert }
{T^\prime _k(\eta ={\pi \over 2})\over  T_k(\eta ={\pi \over 2})}\cdot
{
\int _0^\pi d\sigma ~{\rm P}^{i\zeta }_\np (\cos [\sigma ])~
\sin [k\sigma ]~
\over
\int _0^\pi d\sigma ~{\rm P}^{i\zeta }_\np (\cos [\sigma ])~
\sin [k\sigma ]~/\sin [\sigma ]
}.\eqn\bbc$$
In appendix A we prove that
the right-hand side
of eqn. \bbc ~ is independent of $k$, so that
all of these equations are equivalent.

It readily follows from eqn. \tdep ~ that
$${T^\prime _k(\eta ={\pi \over 2})\over  T_k(\eta ={\pi \over 2})}
=(-2i)\cdot {
\Gamma ({k\over 2}-{\np \over 2}+{1\over 2})
\Gamma ({k\over 2}+{\np \over 2}+1) \over
\Gamma ({k\over 2}-{\np \over 2})
\Gamma ({k\over 2}+{\np \over 2}+{1\over 2})}
\eqn\bbd$$
where we have used $\nu =\np +{1\over 2}.$

We first calculate the integrals in eqn. \bbc ~
for $k=1.$ We rewrite
$$\eqalign{
I_1&=\int _0^\pi d\sigma  ~\sin [\sigma ]~{\rm P}^{i\zeta }_\np (cos[\sigma
])\cr
&=\int _{-\infty }^{+\infty } du ~\sech ^2[u]~
{\rm P}^{i\zeta }_\np (\tanh[ u])\cr
&={1\over \Gamma (1-i\zeta )}\int _{-\infty }^{+\infty }
{du ~e^{i\zeta u}\over \cosh ^2[u]}
{}_2F_1\left( -\np , \np +1; 1-i\zeta ; {1\over 1+e^{2u}}\right) \cr
&={4\over \Gamma (1-i\zeta )}\sum _{n=0}^{+\infty }
{(-\np )_n(\np +1)_n\over (1-i\zeta )_nn!}
\int _{-\infty }^{+\infty }
{du ~e^{i\zeta u}\over e^{-2u}(1+e^{2u})^{n+2}}\cr
&={2\over \Gamma (1-i\zeta )}\sum _{n=0}^{+\infty }
{(-\np )_n(\np +1)_n\over (1-i\zeta )_nn!}
\int _{0}^{+\infty }
{dx ~x^{i\zeta /2}\over (1-x)^{n+2}}\cr
&={2\over \Gamma (1-i\zeta )}\sum _{n=0}^{+\infty }
{(-\np )_n(\np +1)_n\over (1-i\zeta )_nn!}
{\Gamma ({+i\zeta \over 2}+1)\Gamma ({-i\zeta \over 2}+n+1)
\over \Gamma (n+2)}\cr
&={2\Gamma ({+i\zeta \over 2}+1)\Gamma ({-i\zeta \over 2}+1)
\over \Gamma (1-i\zeta )}\cdot
{}_3F_2\left( -\np , \np +1 ,-{i\zeta \over 2}+1;
1-i\zeta , 2;1\right) .\cr }
\eqn\bbe$$
Similarly,
$$\eqalign{
I_2=&\int _0^\pi d\sigma ~{\rm P}^{i\zeta }_\np (cos[\sigma ])\cr
&={2\over \Gamma (1-i\zeta )}\sum _{n=0}^{+\infty }
{(-\np )_n(\np +1)_n\over (1-i\zeta )_nn!}
\int _{-\infty }^{+\infty }
{du ~e^{i\zeta u}\over e^{-u}(1+e^{2u})^{n+1}}\cr
&={1\over \Gamma (1-i\zeta )}\sum _{n=0}^{+\infty }
{(-\np )_n(\np +1)_n\over (1-i\zeta )_nn!}
\int _{0}^{+\infty }
{dx ~x^{i\zeta /2-1/2}\over (1-x)^{n+1}}\cr
&={1\over \Gamma (1-i\zeta )}\sum _{n=0}^{+\infty }
{(-\np )_n(\np +1)_n\over (1-i\zeta )_nn!}
{\Gamma ({+i\zeta \over 2}
+{1\over 2})\Gamma ({-i\zeta \over 2}+n+{1\over 2})
\over \Gamma (n+1)}\cr
&={\Gamma ({+i\zeta \over 2}+{1\over 2})
\Gamma ({-i\zeta \over 2}+{1\over 2})
\over \Gamma (1-i\zeta )}\cdot
{}_3F_2\left( -\np , \np +1 ,-{i\zeta \over 2}+{1\over 2};
1-i\zeta , 1;1\right) .\cr }
\eqn\bbf$$
We simplify the generalized hypergeometric functions using
Whipple's theorem,\refmark{\whipp} which states that
$$
{}_3F_2( a, b, c; e, f; 1)
={\pi \Gamma (e) \Gamma (f)\over
\Gamma ({a+e\over 2}) \Gamma ({a+f\over 2})
\Gamma ({b+e\over 2}) \Gamma ({b+f \over 2})}
\eqn\bbg $$
whenever $a+b=1$ and $e+f=2c+1.$ The generalized
hypergeometric functions
in eqns. \bbe ~ and \bbf ~ satisfy these conditions.

Consequently,
$$\eqalign{
{I_1\over I_2}&=
{2\Gamma ({i\zeta \over 2}+1)\Gamma ({-i\zeta \over 2}+1) \over
\Gamma ({i\zeta \over 2}+{1\over 2})
\Gamma ({-i\zeta \over 2}+{1\over 2})}
{{}_3F_2\left( -\np , \np +1 , -{i\zeta \over 2}+1;
1-i\zeta , 2;1\right) \over
{}_3F_2\left( -\np , \np +1 , -{i\zeta \over 2}+{1\over 2};
1-i\zeta , 1;1\right) } \cr
&={\Gamma ({i\zeta \over 2}+1)\Gamma ({-i\zeta \over 2}+1) \over
\Gamma ({i\zeta \over 2}+{1\over 2})
\Gamma ({-i\zeta \over 2}+{1\over 2})} \cdot
{
\Gamma ({-\np \over 2}+{1\over 2})
\Gamma ({\np \over 2}+1)
\over
\Gamma ({-\np \over 2}+1)
\Gamma ({\np \over 2}+{3\over 2})
},
\cr }\eqn\bbh$$
and
$$\eqalign{\left( {1-c_\zeta \over 1+c_\zeta }
\right) &={i\zeta \over \vert \zeta \vert }
{\Gamma ({+i\zeta \over 2})
\Gamma ({-i\zeta \over 2})\over
\Gamma ({1\over 2}+{+i\zeta \over 2})
\Gamma ({1\over 2}+{-i\zeta \over 2})}
\cr
&={\zeta \over \vert \zeta \vert }\coth [\pi \zeta /2]
\cr
&={1+e^{-\pi \vert \zeta \vert }
\over 1-e^{-\pi \vert \zeta \vert }},\cr }\eqn\bbi$$
so that $c_\zeta =-e^{-\pi \vert \zeta \vert }.$
The result agrees with the previous calculation
for the special case $\np =0$ in ref. [\bgt ].
Surprisingly, the result is independent of $\np .$

Near the null surface the `positive frequency' part
of the inflaton field operator is
$$\eqalign{
&\hat \phi ^{(+)}(u,\tau )={1\over 4\pi }
\int _0^{+\infty }{d\zeta \over
\sqrt {\zeta }}
{\left(e^{\pi \zeta /2}e^{-i\zeta \tau }
-e^{-\pi \zeta /2}e^{+i\zeta \tau }\right) \over
\left(e^{\pi \zeta }-e^{-\pi \zeta }\right) ^{1/2}
}\cdot {1\over \cosh [\tau ]\sech[u]}\cr
&\times \Biggl(
\biggl\{
\sqrt{1+\sqrt{1-\vert C_2(\zeta )\vert ^2/C_1^2(\zeta )}\over 2}
e^{+i\zeta u}\cr
&\hskip 50pt
-e^{i\bar \varphi (\zeta )}
\sqrt{1-\sqrt{1-\vert C_2(\zeta )\vert ^2/C_1^2(\zeta )}\over 2}
e^{-i\zeta u}\biggr\} \cdot
\hat a^{(+)}(+\zeta )\cr
&~~~~+\biggl\{
\sqrt{1+\sqrt{1-\vert C_2(\zeta )\vert ^2/C_1^2(\zeta )}\over 2}
e^{-i\zeta u}\cr
&\hskip 50pt
-e^{-i\bar \varphi (\zeta )}
\sqrt{1-\sqrt{1-\vert C_2(\zeta )\vert ^2/C_1^2(\zeta )}\over 2}
e^{+i\zeta u}\biggr\} \cdot
\hat a^{(+)}(-\zeta )
\Biggr) .\cr }\eqn\bbj$$
where $C_1(\zeta ),$ $C_2(\zeta ),$ and $\bar \varphi (\zeta )$
are defined in eqn. \icla .

\chapter{Continuation into the Open Universe}

In the previous section we expanded the Bunch--Davies
vacuum in region II in terms of the hyperbolic modes.
In this section we continue these modes into region I,
so that we can calculate the power spectrum of Gaussian
adiabatic density perturbations today.

In the new thin wall approximation explained in ref. \bgt,
the effective mass squared of the inflaton field
(equal to $V^{\prime \prime }[\phi _{b}]$
where $\phi _b$ is the background
value for  the inflaton field), changes discontinuously
from $m^2=V^{\prime \prime }[\phi _{fv}]$ to
zero as one passes across the forward light cone
of the materialization center from region II into
region I. This discontinuity is a result of
the approximation, in which we
assume that: (1) the bubble radius (at materialization)
is small compared to the Hubble radius $H^{-1}$ during old
inflation, and (2) the bubble radius (at materialization)
and thickness are small compared to the co-moving wavelengths
of interest. If these two conditions are not satisfied,
the computation becomes more involved, and the results
would then depend on the detailed shape of the potential in
the vicinity of the false vacuum.

In ref. [\bgt ] we derived the following matching
conditions across the light cone
$$\eqalign{
{e^{-i\zeta u}\over \sech [u]}\cdot
{e^{+i\zeta \tau }\over \cosh [\tau ]}
&\to
(+i)\cdot {\sin [\zeta \xi ]\over \sinh[\xi ]}
\cdot e^{(+i\zeta -1)\eta },\cr
{e^{-i\zeta u}\over \sech [u]}\cdot
{e^{-i\zeta \tau }\over \cosh [\tau ]}
&\to 0,\cr
{e^{+i\zeta u}\over \sech [u]} \cdot
{e^{+i\zeta \tau }\over \cosh [\tau ]}
&\to 0,\cr
{e^{+i\zeta u}\over \sech [u]} \cdot
{e^{-i\zeta \tau }\over \cosh [\tau ]}
&\to
(-i)\cdot {\sin [\zeta \xi ]\over \sinh[\xi ]}
\cdot e^{(-i\zeta -1)\eta }.\cr
}\eqn\cca$$
where $\zeta >0.$
The left-hand side indicates asymptotic behavior
in region II as $\sigma \to 0$ [$u\to +\infty $];
the right-hand side indicates asymptotic behavior
in region I as $t\to 0$ [$\eta \to -\infty $].
We define region I conformal time
with the relation $e^\eta =\tanh [t/2].$

For small $t$ the positive frequency part of the
field operator is
$$\eqalign{
\hat \phi ^{(+)}&(\xi ,\eta )=
{(-i)\over 4\pi }\cdot \int _0^{+\infty }{d\zeta \over
\sqrt {\zeta }}
{1\over e^\eta }{\sin [\zeta \xi ]\over \sinh [\xi]}\cr
&\times \Biggl\{ \biggl[
{e^{+\pi \zeta /2} \over \left(e^{\pi \zeta }-e^{-\pi \zeta }\right) ^{1/2}}
\sqrt{1+\sqrt{1-\vert C_2\vert ^2/C_1^2}\over 2}
e^{-i\zeta \eta }\cr
&\hskip 40pt - e^{i\bar \varphi (\zeta )}
{e^{-\pi \zeta /2} \over \left(e^{\pi \zeta }-e^{-\pi \zeta }\right) ^{1/2}}
\sqrt{1-\sqrt{1-\vert C_2\vert ^2/C_1^2}\over 2}
e^{+i\zeta \eta } \biggr] \hat a^{(+)}(\zeta ,+)\cr
& +\biggl[
{e^{-\pi \zeta /2} \over \left(e^{\pi \zeta }-e^{-\pi \zeta }\right) ^{1/2}}
\sqrt{1+\sqrt{1-\vert C_2\vert ^2/C_1^2}\over 2}
e^{+i\zeta \eta }\cr
&\hskip 40pt - e^{-i\bar \varphi (\zeta )}
{e^{+\pi \zeta /2} \over \left(e^{\pi \zeta }-e^{-\pi \zeta }\right) ^{1/2}}
\sqrt{1-\sqrt{1-\vert C_2\vert ^2/C_1^2}\over 2}
e^{-i\zeta \eta } \biggr] \hat a^{(+)}(\zeta ,-)
\Biggr\} .\cr }\eqn\bbbj$$

In ref. [\bgt ] it was shown that the asymptotic behavior
near the null surface for the inflaton field
$\phi \approx e^{\pm i\zeta \eta - \eta }$
corresponds to the asymptotic behavior
$$ \Phi \approx {4\pi GV_{,\phi }\over
(\pm i\zeta +2)}\cdot e^{\pm i\zeta \eta +\eta }
\eqn\dda$$
for the gauge invariant gravitational potential. It
was further shown that the asymptotic behavior for
$\Phi $ above matches onto the exact solution
$$ \Phi = {4\pi GV_{,\phi }\over
(\pm i\zeta +2)}\cdot e^{\pm i\zeta \eta +\eta } \cdot
\left[ 1-{(\zeta \pm i)\over 3(\zeta \mp i)}~e^{2\eta }
\right] \eqn\ddb$$
subject to the following assumptions: (1) $H$ remains
constant, and (2) the potential is linear.
Consequently, to write the positive frequency part
$\Phi ^{(+)},$ we modify eqn. \bbj ~ to become
$$\eqalign{
\hat \Phi ^{(+)}&(\xi ,\eta )=
{(-i)\over 4\pi }{(4\pi GV_{,\phi })\over H}
\cdot e^\eta \cdot \int _0^{+\infty }{d\zeta \over
\sqrt {\zeta }}
{\sin [\zeta \xi ]\over \sinh [\xi]}\cdot
{1\over \left (e^{\pi \zeta }-
e^{-\pi \zeta }\right) ^{1/2}}\cr
&\times \Biggl\{ \biggl[
e^{+\pi \zeta /2}
\sqrt{1+\sqrt{1-\vert C_2\vert ^2/C_1^2}\over 2}
{e^{-i\zeta \eta }\over 2-i\zeta }
\left\{  1-{(\zeta - i)\over 3(\zeta + i)}~e^{2\eta } \right\}  \cr
&\hskip 10pt - e^{i\bar \varphi (\zeta )}
e^{-\pi \zeta /2}
\sqrt{1-\sqrt{1-\vert C_2\vert ^2/C_1^2}\over 2}
{e^{+i\zeta \eta }\over 2+i\zeta }
\left\{  1-{(\zeta + i)\over 3(\zeta - i)}~e^{2\eta }
\right\}   \biggr] \hat a^{(+)}(+\zeta )\cr
& +\biggl[
e^{-\pi \zeta /2}
\sqrt{1+\sqrt{1-\vert C_2\vert ^2/C_1^2}\over 2}
{e^{+i\zeta \eta }\over 2+i\zeta }
\left\{  1-{(\zeta + i)\over 3(\zeta - i)}~e^{2\eta } \right\}   \cr
&\hskip 10pt - e^{-i\bar \varphi (\zeta )}
e^{+\pi \zeta /2}
\sqrt{1-\sqrt{1-\vert C_2\vert ^2/C_1^2}\over 2}
{e^{-i\zeta \eta }\over 2-i\zeta }
\left\{  1-{(\zeta - i)\over 3(\zeta + i)}~e^{2\eta } \right\} \biggr]
\hat a^{(+)}(-\zeta )
\Biggr\} .\cr }\eqn\ddc$$
Even though we are working in units with $H=1,$ we insert the $H^{-1}$
factor to facilitate later conversion to more conventional units in
the final result.
We define the power spectrum for $\Phi $
according to the following relation for the
two point function:
$$ \langle \Phi (\xi =0, t)
\Phi (\xi , t)\rangle =\int _0^\infty d\zeta \cdot \zeta ~
{\sin [\zeta \xi ]\over \sinh [\xi ]}
P_{\Phi }(\zeta, t) .\eqn\ddd $$

Taking the limit $t\to \infty $ [$\eta \to 0^-$]
isolates the growing mode. $P_{\Phi }$ shall
denote the limit of $P_{\Phi }(\zeta,t)$ as $t\to \infty .$
It follows from eqn. \ddc ~ that
$$\eqalign{&P_\Phi (\zeta )=
(GV_{,\phi })^2\cdot {4\over 9H^2}\cdot {1\over \zeta (\zeta ^2+1)}
\cdot {1\over \left(e^{\pi \zeta }-
e^{-\pi \zeta }\right) } \cr
&\times \Biggr[ ~\Biggl|
e^{+\pi \zeta /2}
\sqrt{1+\sqrt{1-\vert C_2\vert ^2/C_1^2}\over 2}
+e^{+i\bar \varphi (\zeta )} \cdot \left( {\zeta +i\over \zeta -i}\right)
e^{-\pi \zeta /2}
\sqrt{1-\sqrt{1-\vert C_2\vert ^2/C_1^2}\over 2}
\Biggr| ^2\cr
&+ \Biggl|
e^{-\pi \zeta /2}
\sqrt{1+\sqrt{1-\vert C_2\vert ^2/C_1^2}\over 2}
+e^{-i\bar \varphi (\zeta )} \cdot \left( {\zeta -i\over \zeta +i}\right)
e^{+\pi \zeta /2}
\sqrt{1-\sqrt{1-\vert C_2\vert ^2/C_1^2}\over 2}
\Biggr| ^2\Biggr] \cr
&=(GV_{,\phi })^2\cdot {4\over 9}\cdot {1\over \zeta (\zeta ^2+1)}
\cdot {1\over \left(e^{\pi \zeta }-
e^{-\pi \zeta }\right) } \cr
& \times \left[ e^{\pi \zeta }+e^{-\pi \zeta }+
{\vert C_2\vert \over C_1}
\cdot \left( e^{i\bar \varphi }\cdot {\zeta +i\over
\zeta -i} + e^{-i\bar \varphi }\cdot {\zeta -i\over
\zeta +i}\right) \right ] .\cr } \eqn\dde$$

Using $\chi =16\pi G(V^2/V_{,\phi }^2)
\Phi $ (during inflation), one obtains that the
power spectrum for $\chi $ (which has a normalization
that relates more directly to the density perturbations
seen after reheating) is
$$\eqalign{&P_\chi (\zeta )=
\left[ 16\pi G\left( {V^2\over V_{,\phi }^2}\right) \right] ^2
P_\Phi (\zeta )\cr
&={9\over 4\pi ^2}\cdot \left({H^3\over V_{,\phi }}\right) ^2\cdot {1\over
\zeta (\zeta ^2+1)}
\cdot \left[ {
 e^{\pi \zeta }+e^{-\pi \zeta }+
{\vert C_2\vert \over C_1}
\cdot \left( e^{i\bar \varphi }\cdot {\zeta +i\over
\zeta -i} + e^{-i\bar \varphi }\cdot {\zeta -i\over
\zeta +i}\right) \over \left(e^{\pi \zeta }-
e^{-\pi \zeta }\right) }\right ] \cr }
\eqn\pspchi$$
where
$$\chi ={2\over 3}{H^{-1}\dot \Phi +\Phi \over 1+w}+\Phi .
\eqn\defchi$$
Computing the power spectrum in the limit $t\to +\infty ,$
which formally assumes inflation without end, has the effect
of completely eliminating the decaying mode.
While $\Omega $ is close to one (which is true here
except for the very early part of the new inflationary epoch)
the variable $\chi $ is conserved on superhorizon scales
irrespective of changes in $H$ or in the slope of the potential.
Note with the conventions used here $P\sim \zeta ^{-3}$ corresponds
to scale invariance.

Our result differs from that of Lyth and Stewart\refmark{\lyth }
and of Ratra and Peebles \refmark{\ratra, \ratrb }
(who assume different initial conditions for the quantum fields)
only by the factor in square brackets in eqn. \pspchi .
Because ${\vert C_2\vert /C_1}<1$,
the bracketed quantity lies between $\tanh [\pi \zeta /2]$
and  $\coth [\pi \zeta /2].$ This fact severely limits the
influence of mass at large wave numbers.
Observations of the large angle CMB anisotropy, for example,
are only sensitive to values of $\zeta > 1$ or so.
We conclude that the idea of open inflation is
actually more predictive than it might appear at first
sight. In Fig. 3 we plot the bracketed quantity versus
co-moving wavenumber for various values of $m^2/H^2.$
The envelope of two dotted curves indicates the bounds
$\tanh [\pi \zeta /2]$ and $\coth[ \pi \zeta /2].$

The phenomenology of the Ratra-Peebles spectrum has
been explored in a number of recent papers, with
the assumption that the inflaton potential $V(\phi)$
is linear. The conclusion of those papers was that
for $\Omega_0 \sim 0.3-0.4$, adiabatic density perturbations
of the form studied here were consistent with most
current observational constraints.\refmark{\kamion ,\gorski }

We should point out, however, that in our scenario,
which provides physically motivated initial
conditions, there is no very strong reason to
expect the potential to be linear over the range of $\phi$
of interest. By shortening the length of the
`slow-roll' transition and assuming that the potential
{\it does } have significant structure (i.e. a
false vacuum) for the relevant range of $\phi$,
we are increasing the sensitivity of
the final  perturbation spectrum to the details
of the potential. For example, it may be that
within our framework a positively tilted spectrum
(i.e., increasing power at shorter wavelengths)
is quite likely, and this could restore the viability
of models with even lower values of $\Omega_0$.

\chapter{Concluding Remarks}

We conclude with the following comments:

1. In  ref. [\bgt ] we calculated the power spectrum
for the `conformal' mass case $m^2/H^2=2.$ There
was no physical motivation for choosing this mass;
the choice was made solely for computational
simplicity. Naively one might expect that allowing
$m^2/H^2$ to be a free parameter would diminish
the predictiveness of open inflation. However,
this does not turn out to be the case, except
for very small values of $\Omega _0,$ which
can be ruled out observationally based on
lower bounds on the mean mass density of the
universe. For scales smaller than the curvature
scale there is little freedom to alter the
density perturbations by adjusting $m^2/H^2.$

2. The assumptions made here to calculate the
power spectrum should be stressed. We assumed that at
materialization the size of the bubble and the bubble
wall thickness are both small compared to the Hubble
radius $H^{-1}$ during {\it old} inflation.
Relaxing
these assumptions would alter the power spectrum
calculated here. If at materialization the bubble covers
an appreciable fraction of the Hubble volume during old
inflation, a precise calculation of the perturbations
requires taking into account the effect of the Euclidean
(classically forbidden) evolution of the background
solution on the evolution of the inflaton field perturbations.
Alternatively, if one considers perturbations on
a co-moving scale not large compared to the bubble
wall thickness (or bubble radius at materialization), details
of the bubble wall profile, which are highly model dependent,
become relevant. Our discontinuous treatment of the mass
change across the bubble wall may be regarded as a sudden
impulse approximation, which breaks down for sufficiently
large co-moving wave numbers.

In region I we assumed that the potential was exactly
linear and that over the range of interest $V[\phi ]$ does
not change appreciably, so that $H$ could be regarded as
constant. These are precisely the same assumptions that
give an {\it exactly} scale invariant power spectrum for
standard $(\Omega _0=1)$ new inflation. In the
recent literature, it has
been stressed that because of the change in $H$
during inflation and because of variation in the
slope of the potential during inflation, deviations
from exact scale invariance are to be expected. These
considerations apply equally well to open inflation.
The power spectrum calculated here should be regarded
as the small--$\Omega _0$ analogue of exact scale invariance
for the flat $(\Omega _0=1)$ case. It is the generic
prediction from which small deviations that depend
on the exact choice of potential are to be expected.

As mentioned at the end of the last section,
in the open case there may be more reason to
expect a tilted spectrum than in the flat case,
because in open inflation the epoch of new inflation
begins near a local maximum of the potential.

3. Finally, for completeness it should be noted that
when $0\le m^2/H^2<2,$ there exists an additional
`bound state' mode in region II (for the s-wave).
This can readily be seen by rewriting eqn. \icc ~
in terms of $F$ and $u$ as
$${\partial ^2F\over \partial u^2}+
\biggl[ \zeta ^2 +\left( {m^2\over H^2}-2\right)
\sech ^2[u]\biggr] F(u)=0.\eqn\www$$
When $0\le m^2/H^2<2,$ there exists a single bound
state, which must be included into the mode expansion
in region II for completeness. This mode, however,
does not propagate into region I, because as
$u \to \infty $ (or alternatively as $\sigma \to 0$)
and one approaches the boundary of region I, this
mode decays exponentially in $u.$ It is a mode
confined to region II
and thus can safely be ignored for the calculation
of the density perturbations inside the bubble.

4. For small values of $m^2/H^2$ one has to worry
about whether the tunneling is described by an
$SO(3,1)$ symmetric bounce (i.e., the Coleman--de
Luccia instanton). When $\vert V^{\prime \prime }\vert
/H^2<4$ at the top of the barrier (i.e., at the local
maximum), the tunneling is described by the Hawking--Moss
instanton\refmark{\hmoss}  rather than the
Coleman--de Luccia instanton.\refmark{\jensen }
In this regime there is no expanding bubble solution
that is $SO(3,1)$ symmetric, and it seems doubtful
that a sufficiently
homogeneous and isotropic universe will result. Moreover,
in the limit as the Coleman--de Luccia instanton approaches
the Hawking--Moss instanton, the power spectrum on large
scales diverges, because of the flatness of the potential
near the local maximum. The physical interpretation of the
Hawking--Moss instanton has been discussed in refs.
[\newa ]. Generally, one would expect $m^2/H^2$ (at the
local minimum) to be comparable to $\vert V^{\prime \prime }\vert
/H^2$
at the local maximum, so it would appear difficult to
construct viable models with very small $m^2/H^2.$

\centerline{***************************}

The implications of the power spectra calculated here
for the cosmic microwave background will be discussed
elsewhere.\refmark{\bucher}

{\bf Note Added:} After this work was completed, we received
a preprint by K. Yamamoto, M. Sasaki, and T. Tanaka
on the CMB anisotropy from open inflation.\refmark{\yst }
Their results do not agree with ours. Several differences
should be noted:

(1). Although the same scenario that we
considered is described in sections I and II
of ref. [\yst ], in section III they apparently switch
scenarios to one in which the mass of the scalar field
does not change across the bubble wall. [They
assume that $m^2/H^2\ll 1$ everywhere, which is
problematic. (See Note 4 above.)]

(2). In ref. [\yst ] it is
claimed that additional discrete modes must be included.
In region I including extra discrete modes is
not necessary and leads to an overcomplete set of modes.
Although for $0\le m^2/H^2<2$ there
are additional discrete modes in region II,
these do not propagate into region I. (See
Note 3 above.)

(3). The state of the scalar field modes is not determined by
matching from an earlier epoch of {\it old} inflation, as we
have done. Instead it is assumed that the correct modes can be
obtained by analytically continuing the `de Sitter invariant
Euclidean vacuum' modes {\it inside} the bubble, in effect
assuming that the bubble wall leaves no imprint on the scalar
field fluctuations. Our calculation indicates this assumption
to be incorrect.

{\bf Acknowledgements:} We would like to thank A.S. Goldhaber,
A. Linde, B. Ratra, M. Sasaki, and K. Yamamoto
for useful discussions. This work was partially supported by
NSF contract PHY90-21984 and by the David and Lucile Packard
Foundation.

\refout

\centerline{\bf Figure Captions}

\item{Fig. 1}{ A diagram of maximally extended
de Sitter space, divided into the five hyperbolic coordinate
patches. The vertical dashed lines indicate radial coordinate
singularities. $M$ is the materialization center and $\bar M$
is its antipodal point. The forward and backward light cones
of $M$ contain regions I and V, and the forward and backward
light cones of $\bar M$ contain regions III and IV, respectively.
Note that of the five regions, only region II contains a
complete Cauchy surface for all of de Sitter space.}

\item{Fig. 2}{Bubble Nucleation Process.}

\item{Fig. 3}{Dependence of the Power spectrum on $m^2/H^2.$
The bracketed quantity in eqn. \pspchi ~ is plotted as a function
of co-moving wavenumber $\zeta $ for various values of
$m^2/H^2$ (solid curve). The dashed envelope indicates the lower
and upper bounds $\tanh [\pi \zeta /2]$ and $\coth [\pi \zeta /2]$
for this quantity.}

\APPENDIX{A}{}

In this appendix we verify that the right-hand side of
eqn. \bbc ~ is independent of $k.$
Using eqn. \bbd , and the substitution $x= \cos [\sigma] = \tanh[u]$,
$d \sigma = \sin[\sigma] d\sigma$, we need to show that
$$
{\Gamma ({k\over 2}-{\np \over 2}+{1\over 2})
\Gamma ({k\over 2}+{\np \over 2}+1) \over
\Gamma ({k\over 2}-{\np \over 2})
\Gamma ({k\over 2}+{\np \over 2}+{1\over 2})}
{\int_{-\infty}^\infty du ~{\rm P}^{i\zeta }_\np (\cos [\sigma ])~
\sin [k\sigma ]~ \sin [\sigma]
\over
\int_{-\infty}^\infty  du ~{\rm P}^{i\zeta }_\np (\cos [\sigma ])~
\sin [k\sigma ]
}\eqn\abbc$$
is independent of $k$. We denote the integral in the numerator
$A_{\np,k}$ and that in the denominator $B_{\np,k}$.

{}From the standard recursion relation for Legendre functions
(e.g., ref.  [\erdelyi ],
p. 1005),
$$
(1+\np -i\zeta) ~{\rm P}^{i\zeta }_{\np+1} (x)~ = (\np+1) x ~{\rm P}^{i\zeta
}_\np(x) +{d ~\over du}~{\rm P}^{i\zeta }_\np(x),
\eqn\abbd
$$
it follows upon substitution into the integrals,
integrating by parts, and using trigonometric double
angle formulae that
$$
\eqalign{
A_{\np,k+1}(\np-k) &= 2(1+\np-i\zeta) A_{\np+1,k} -(\np+k) A_{\np,k-1}\cr
B_{\np,k+1}(1+\np-k) &= 2(1+\np-i\zeta) B_{\np+1,k} -(1+\np+k) B_{\np,k-1}.\cr
}
\eqn\abbe
$$
which, since $A_{\np,0}=B_{\np,0}=0$, give all the integrals
we need in terms of those for $k=1$.
Redefining
$$
\eqalign{
C_{\np,k}& =
\Gamma ({k\over 2}-{\np \over 2}+{1\over 2})
\Gamma ({k\over 2}+{\np \over 2}+1) A_{\np,k}\cr
D_{\np,k} &=\Gamma ({k\over 2}-{\np \over 2})
\Gamma ({k\over 2}+{\np \over 2}+{1\over 2}) B_{\np,k},\cr
}
\eqn\abbf
$$
(which are well defined for arbitrary positive mass)
we find from eqn. \abbe ~
that $C$ and $D$ obey identical recursion relations:
$$
\eqalign{
C_{\np,k+1}& =
- (1+\np-i\zeta) C_{\np+1,k} + {1\over 4} (k+\np ) (1+k+\np) C_{\np,k-1},\cr
D_{\np,k+1}& =
- (1+\np-i\zeta) D_{\np+1,k} + {1\over 4} (k+\np ) (1+k+\np) D_{\np,k-1}.\cr
}
\eqn\abbg
$$
Since both vanish for $k=0$,
it follows that $C_{\np,k}/ D_{\np,k} = C_{\np,1}/ D_{\np,1}$ and
thus that
eqn. \abbc ~ is independent of $k$, as wanted.

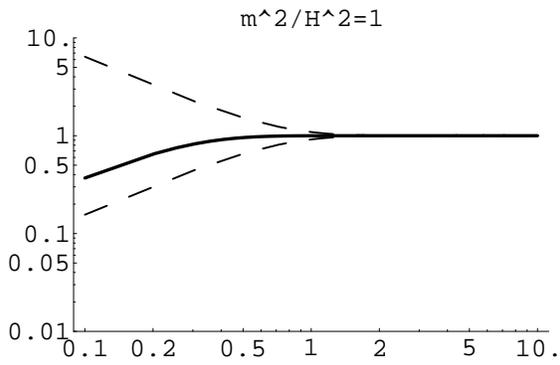
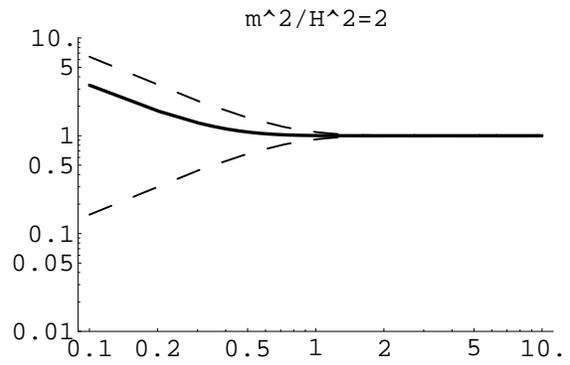
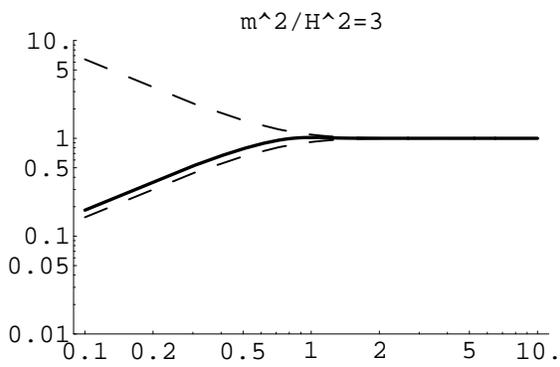
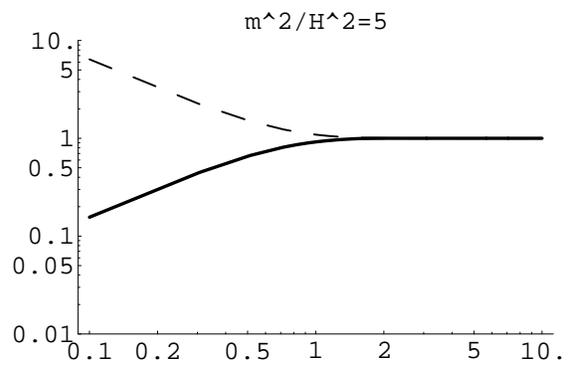
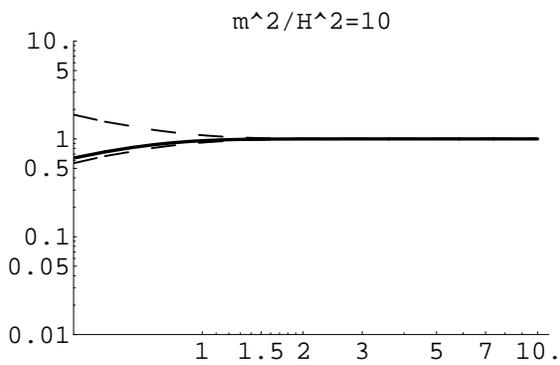
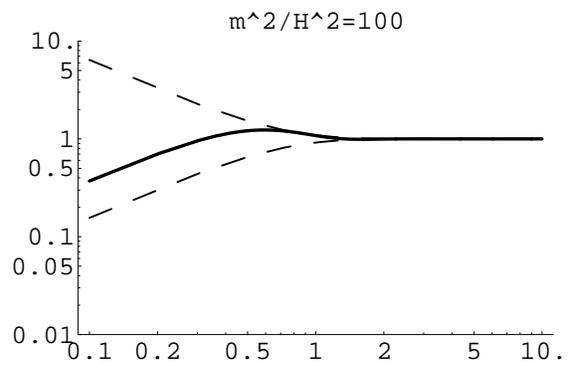
\end